\documentstyle[12pt]{article}
\begin{document}
\newcommand{\siml}{\stackrel{<}{\sim}}
\newcommand{\simg}{\stackrel{>}{\sim}}
\newcommand{\lleq}{\stackrel{<}{=}}

\baselineskip=1.333\baselineskip


%
\begin{center}
{\large\bf
Effects of multiplicative noises on
synchronization 
in finite $N$-unit stochastic ensembles:
Augmented moment approach
} 
\end{center}

\begin{center}
Hideo Hasegawa
\footnote{e-mail:  hasegawa@u-gakugei.ac.jp}
\end{center}

\begin{center}
{\it Department of Physics, Tokyo Gakugei University  \\
Koganei, Tokyo 184-8501, Japan}
\end{center}
\begin{center}
({\today})
\end{center}
\thispagestyle{myheadings}

\begin{abstract}
We have studied the synchronization in 
finite $N$-unit FitzHugh-Nagumo
neuron ensembles 
subjected to additive and multiplicative noises,
by using the augmented moment method (AMM) which is
reformulated with the use of the Fokker-Planck equation.
It has been shown that for diffusive couplings,
the synchronization may be enhanced
by multiplicative noises while additive noises are detrimental to
the synchronization.
In contrast, for sigmoid coupling,
both additive and multiplicative noises
deteriorate the synchronization.
The synchronization depends not only on the type of noises
but also on the kind of couplings.

\end{abstract}

\vspace{0.5cm}

%
\newpage
\section{INTRODUCTION}


Nonlinear stochastic equations subjected 
to additive and/or multiplicative noises
have been widely adopted for a study
on real systems in physics,
biology, chemistry, economy and networks.
Interesting phenomena caused by both the noises have
been intensively investigated
(for a recent review, see Ref. 1, related references therein).
It has been realized that
the property of multiplicative noises
is different from that of additive noises
in some respects as follows.
(1) Multiplicative noises induce the phase transition,
creating an ordered state, while additive noises
are against the ordering \cite{Broeck94}-\cite{Munoz05}.
(2) Although the stochastic resonance is not realized
in linear systems with additive noises,
it may be possible with multiplicative {\it color} noise
(but not with multiplicative {\it white} noise) 
\cite{Berd96,Bazykin97}.
(3) Although the probability distribution in stochastic systems
subjected to additive Gaussian noise follows the Gaussian,
it is not the case for multiplicative
Gaussian noises which generally yield non-Gaussian distribution
\cite{Sakaguchi01}-\cite{Hasegawa05b}.
(4) The scaling relation of the effective 
strength for additive noise given by
$\beta(N)=\beta(1)/\sqrt{N}$ is not applicable to
that for multiplicative noise:
$\alpha(N) \neq \alpha(1)/\sqrt{N}$, where $\alpha(N)$ and $\beta(N)$
denote effective strengths of multiplicative
and additive noises, respectively, in the $N$-unit system
\cite{Hasegawa06}.

In order to show the above item (4), the present author has
adopted the augmented moment method (AMM) in a recent paper
\cite{Hasegawa06}.
The AMM was originally
developed by expanding variables
around their mean values in order to obtain
the second-order moments both for
local and global variables in stochastic systems 
\cite{Hasegawa03a}.  
The AMM has been successfully
applied to a study on dynamics of coupled stochastic systems 
described by Langevin, FitzHugh-Nagumo and
Hodgkin-Huxley models subjected only to additive noises
with global, local or small-world couplings
(with and without transmission delays)
\cite{Hasegawa}. 
In Ref. \cite{Hasegawa06}, we have reformulated the AMM  
with the use of the Fokker-Planck equation (FPE),
in order to avoid the difficulty
due to the Ito versus Stratonovich calculus inherent
for multiplicative noise. 
It has been pointed our that a naive approximation
of the scaling relation for multiplicative noise: 
$\alpha(N)=\alpha(1)/\sqrt{N}$, 
as adopted by Mu$\tilde{\rm n}$oz, Colaiori
and Castellano in their recent paper \cite{Munoz05},
leads to the result
which violates the central-limit theorem and
which is in disagreement with
those of AMM and direct simulations.

The purpose of 
the present paper is two folds:
(1) to reformulate AMM forFitzHugh-Nagumo (FN) model

subjected to both additive and multiplicative noises
with the use of FPE \cite{Hasegawa06}, 
and (2) to discuss the respective roles of the two noises 
on the synchronization. 
Our calculations have shown that multiplicative noises may
enhance the synchronization while additive noises
work to destroy it.
This is similar to the property
in item (1) discussed above.

The paper is organized as follows.
In Sec. II, we have applied the DMA
to finite $N$-unit FN networks subjected to
additive and multiplicative noises.
Numerical calculations are presented in Sec. III.
The final Sec. IV is devoted to conclusion and discussion. 

\section{Noisy FN neuron ensembles}

\subsection{Augmented moment method}

We have adopted $N$-unit FN neurons
subjected to additive and multiplicative noises.
Dynamics of a neuron $i$ in a given FN neuron ensemble
is described by the nonlinear differential equations
(DEs) given by 
\begin{eqnarray}
\frac{dx_{i}}{dt} &=& F(x_{i})
- c \:y_{i}
+\alpha\: G(x_i) \eta_i(t)+ \beta \:\xi_i(t)
+I_i^{(c)}(t)+I^{(e)}(t), \\
\frac{dy_{i}}{dt} &=& b \:x_{i} - d \:y_{i}+e,
\hspace{2cm}\mbox{($i=1$ to $N$)}
\end{eqnarray}
with
\begin{equation}
I_i^{(c)}(t)= \frac{J}{Z} \sum_{j\neq i} (x_j-x_i).
\end{equation}
In Eq. (1)-(3),
$F(x)=k x (x-a) (1-x)$, 
$k=0.5$, $a=0.1$, $b=0.015$, $d=0.003$ and $e=0$
\cite{Hasegawa03a}\cite{Rod96}: 
$x_i$ and $y_i$ denote the fast (voltage) variable
and slow (recovery) variable, respectively:
$G(x)$ an arbitrary function of $x$:
$I^{(e)}(t)$ an external input whose explicit form
will be shown shortly [Eq. (41)]:
$J$ expresses strengths of diffusive couplings, $Z=N-1$: 
$\alpha$ and $\beta$ denote magnitudes of
multiplicative and additive noises, respectively,
and $\eta_i(t)$ and $\xi_i(t)$ express zero-mean Gaussian white
noises with correlations given by
\begin{eqnarray}
\langle \eta_i(t)\:\eta_j(t') \rangle
&=& \delta_{ij} \delta(t-t'),\\
\langle \xi_i(t)\:\xi_j(t') \rangle 
&=& \delta_{ij} \delta(t-t'),\\
\langle \eta_i(t)\:\xi_j(t') \rangle &=& 0.
\end{eqnarray}
  
The Fokker-Planck equation $p(\{x_i\},\{y_i\},t)$
is expressed 
in the Stratonovich representation by\cite{Hasegawa06}\cite{Haken83}
\begin{eqnarray}
\frac{\partial}{\partial t} p
&=&-\sum_k \frac{\partial}{\partial x_k}\{ [F(x_k)-cy_k+I_k]p \}
- \sum_k \frac{\partial}{\partial y_k}[(bx_k-dy_k+e)p] \nonumber \\
&&+\frac{\alpha^2}{2}\sum_k \frac{\partial}{\partial x_k}
\{ G(x_k)\:\frac{\partial}{\partial x_k}[G(x_k)\:p] \}
+\sum_k \frac{\beta^2}{2}\frac{\partial^2 }{\partial x_k^2}\:p,
\end{eqnarray}
where $I_k=I_k^{(c)}+I^{(e)}$.

We are interested also in dynamics of
global variables $X(t)$ and $Y(t)$ defined by
\begin{eqnarray}
X(t)=\frac{1}{N} \sum_i x_i(t), \\
Y(t)=\frac{1}{N} \sum_i y_i(t). 
\end{eqnarray}
The probability of $P(X,Y,t)$ is expressed 
in terms of $p(\{x_i\},\{y_i\},t)$ by
\begin{equation}
P(X,Y,t)= \int \int dx_i dy_i\: \Pi_i \:p_i(x_i,y_i,t)
\:\delta(X-\frac{1}{N} \sum_i x_i) 
\:\delta(Y-\frac{1}{N} \sum_i y_i).
\end{equation}
Moments of local and global variables
are expressed by
\begin{eqnarray}
\langle x_i^k \:y_i^{\ell} \rangle
&=&
= \int \int d x_i d y_y \;p_i(x_i,y_i,t) \:x_i^k y_i^{\ell}, \\
\langle X^k\:Y^{\ell} \rangle
&=& \int \int dX dY P(X,Y,t) \;X^k Y^{\ell}.
\end{eqnarray}
By using Eqs. (1), (2), (7) and (11), 
we get equations of motions for
means, variances and covariances of local variables by
\begin{eqnarray}
\frac{d \langle x_i \rangle}{dt}&=& \langle F(x_i) \rangle 
-c \langle y_i \rangle
+\frac{\alpha^2}{2} \langle G'(x_i)G(x_i) \rangle, \\
\frac{d \langle y_i \rangle }{dt}
&=&b \langle x_i \rangle -d  \langle y_i \rangle +e, \\
\frac{d \langle x_i x_j \rangle}{dt}
&=& \langle x_i F(x_j) \rangle 
+  \langle x_j F(x_i) \rangle 
-c(\langle x_i y_j \rangle + \langle x_j y_i \rangle)
\nonumber \\
&&+\frac{J}{Z}\sum_k (\langle x_ix_k \rangle
+\langle x_jx_k \rangle  - \langle x_i^2\rangle
-\langle x_j^2 \rangle) \nonumber \\
&&+\frac{\alpha^2}{2} [\langle x_iG'(x_j)G(x_j) \rangle 
+\langle x_jG'(x_i)G(x_i) \rangle ] \nonumber \\
&&+[\alpha^2 \langle G(x_i)^2 \rangle +\beta^2] \delta_{ij}, \\
\frac{d \langle y_i y_j \rangle }{dt}
&=&b (\langle x_iy_j \rangle + \langle x_jy_i \rangle)
-2 d \langle y_i y_j \rangle, \\
\frac{d \langle x_i\:y_j \rangle}{dt}
&=& \langle y_j\:F(x_i) \rangle 
-c \langle y_i y_j \rangle +b \langle x_ix_j \rangle
-d \langle x_i\:y_j \rangle  \nonumber \\
&&+\frac{w}{Z}\sum_k ( \langle x_k\:y_j \rangle -\langle x_i\:y_j \rangle) 
+\frac{\alpha^2}{2} \langle y_j\:G'(x_i)G(x_i) \rangle, 
\end{eqnarray}
where $G'(x)=d G(x)/d x$.

Equations of motions
for variances and covariances of global variables 
are obtainable from Eqs. (8), (9) and (12):
\begin{eqnarray}
\frac{d \langle V_{\kappa} \rangle}{dt}
&=& \frac{1}{N} \sum_i \langle v_{\kappa i} \rangle , \\
\frac{d \langle V_{\kappa}\:V_{\lambda} \rangle}{dt}
&=& \frac{1}{N^2}\sum_i\sum_j 
\frac{d \langle v_{\kappa i}\:v_{\lambda j} \rangle}{dt},
\hspace{1cm}\mbox{($\kappa, \gamma=1,2$)}
\end{eqnarray}
where we adopt a convention: 
$v_{1i}=x_i$, $v_{2i}=y_i$, $V_1=X$ and $V_2=Y$.
Equations (13) and (14) are used for $N=1$ FN neuron ($\alpha=0$)
and for $N=\infty$ FN neuron ensembles ($\alpha=0$)
in the mean-field approximation \cite{Aceb04}.
Equations (13)-(17) are employed
in the moment method for a single FN neuron
subjected to additive noises \cite{Rod96}. 
We will show that Eqs. (18) and (19) play important roles
in discussing finite $N$-unit FN ensembles.

In the AMM \cite{Hasegawa03a}, we define the eight quantities given by
\begin{eqnarray}
\mu_{\kappa} &=& \langle V_{\kappa} \rangle
= \frac{1}{N} \sum_i \langle v_{\kappa i} \rangle, \\
\gamma_{\kappa,\lambda}
&=&\frac{1}{N} \sum_i \langle (v_{\kappa i}-\mu_{\kappa})
\:(v_{\lambda i}-\mu_{\lambda}) \rangle, \\
\rho_{\kappa,\lambda}
&=& \langle (V_{\kappa}-\mu_{\kappa})\:(V_{\lambda}-\mu_{\lambda}) \rangle, 
\hspace{1cm}\mbox{($\kappa, \gamma=1,2$)}
\end{eqnarray}
with $\gamma_{1,2}=\gamma_{2,1}$ and $\rho_{1,2}=\rho_{2,1}$.
It is noted that $\gamma_{\kappa,\lambda}$ 
expresses the averaged
fluctuations in local variables while
$\rho_{\kappa,\lambda}$ denotes fluctuations in global variables.
Expanding Eqs. (13)-(19) around means of 
$\mu_{\kappa}$ as $v_{\kappa i}=\mu_{\kappa}+\delta v_{\kappa i}$,
we get equations of motions for the eight quantities:
\begin{eqnarray}
\frac{d \mu_1}{dt} &=& f_o+f_2\gamma_{1,1}-c \mu_2
+\frac{\alpha^2 \mu_1}{2}+I^{(e)}, \\
\frac{d \mu_2}{dt}&=& b \mu_1-d \mu_2+e, \\
\frac{d \gamma_{1,1}}{dt} &=& 2(a \gamma_{1,1}-c \gamma_{1,2})
+\frac{2J N}{Z} (\rho_{1,1}-\gamma_{1,1})
+2 \alpha^2 \gamma_{1,1} + \alpha^2 \mu_1^2
+\beta^2, \\
\frac{d \gamma_{2,2}}{dt}&=& 2(b\gamma_{1,2}-d\gamma_{2,2}), \\
\frac{d \gamma_{1,2}}{dt} &=& b\gamma_{1,1}
+(a-d)\gamma_{1,2}-c\gamma_{2,2}
+\frac{J N}{Z}(\rho_{1,2}-\gamma_{1,2})
+\frac{\alpha^2\gamma_{1,2}}{2}, \\
\frac{d \rho_{1,1}}{dt} &=& 2(a \rho_{1,1}-c \rho_{1,2})
+2 \alpha^2 \rho_{1,1} + \frac{\alpha^2 \mu_1^2}{N} 
+\frac{\beta^2}{N}, \\
\frac{d \rho_{2,2}}{dt}&=& 2(b\rho_{1,2}-d\rho_{2,2}), \\
\frac{d \rho_{1,2}}{dt} &=& b\rho_{1,1}
+(a-d)\rho_{1,2}-c\rho_{2,2}
+\frac{\alpha^2\rho_{1,2}}{2}, 
\end{eqnarray}
where $a=f_1+3 f_3 \gamma_{1,1}$,
$f_{\ell}=(1/\ell \:!)F^{(\ell)}(\mu_1)$, 
and $G(x)=x$ is adopted,
relevant expressions for a general $G(x)$ being given in the appendix.
The original $2N$-dimensional stochastic equations
given by Eqs. (1) and (2)
are transformed to eight-dimensional deterministic
equations. Equations (23)-(30)
with $\alpha=0$ (additive noises only)
reduce to those obtained previously 
\cite{Hasegawa03a}.

\subsection{$N$ dependence of effective noise strength}

Comparing the $\beta^2$ term in $d \gamma_{1,1}/dt$  
of Eq. (25) to that in $d \rho_{1,1}/dt$ of Eq. (28),
we note that the effective strength of additive noise
is scaled by 
\begin{equation}
\beta \rightarrow \frac{\beta}{\sqrt{N}}.
\end{equation}
As for multiplicative noise, however, the situation is
not so simple.
A comparison between the $\alpha^2$ terms in Eq. (27) and (30)
yield the two kinds of scalings:
\begin{eqnarray}
\alpha &\rightarrow& \frac{\alpha}{\sqrt{N}},
\hspace{1cm}\mbox{for $\mu_1$ term}, \\
\alpha &\rightarrow& \alpha,
\hspace{1cm}\mbox{for $\gamma_{1,1}$ and $\rho_{1,1}$ terms}, 
\end{eqnarray}
The relations given by Eqs. (31)-(33) hold
also for $d \gamma_{1,2}/dt$ and $d \rho_{1,2}/dt$
given by Eqs. (27) and (30).
Thus the scaling behavior of the effective strength of
multiplicative noises is quite different
from that of additive noises, as previously pointed out
for Langevin model \cite{Hasegawa06}.

Nevertheless, we note that in the limit of $J=0$,
AMM equations lead to
\begin{eqnarray}
\rho_{\kappa,\lambda} &=& \frac{\gamma_{\kappa,\lambda}}{N},
\hspace{1cm}\mbox{($\kappa, \lambda=1,2)$}
\end{eqnarray}
which is nothing but the central-limit theorem
describing the relation between fluctuations
in local and average variables.

\subsection{Synchronization ratio}

In order to quantitatively discuss the
synchronization, we first consider the quantity given by
\cite{Hasegawa03a}
\begin{equation}
R(t)=\frac{1}{N^2} \sum_{i j} \langle [x_i(t)-x_j(t)]^2 \rangle
=2 [\gamma_{1,1}(t)-\rho_{1,1}(t)].
\end{equation}
When all neurons are in the completely synchronous state,
we get $x_{i}(t)=X(t)$ for all $i$, and 
then $R(t)=0$ in Eq. (35).
On the contrary, in the asynchronous state, we get 
$R(t)=2(1-1/N)\gamma_{1,1} \equiv R_0(t)$
from Eq. (34).
We have defined the synchronization ratio
given by \cite{Hasegawa03a}
\begin{equation}
S(t) =1-\frac{R(t)}{R_0(t)}
= \left( \frac{N\rho_{1,1}(t)/\gamma_{1,1}(t)-1}{N-1} \right),
\end{equation}
which is 0 and 1 for completely asynchronous ($R=R_0$)  
and synchronous states ($R=0$), respectively.
We have studied the synchronization ratios 
at $t_f$ and $t_m$ as given by
\begin{eqnarray}
S_{f}&=&S(t_{f}), \\
S_{m}&=&S(t_{m}),
\end{eqnarray}
with
\begin{eqnarray}
t_{f}&=&\{t \mid X(t)=\theta, d X(t)/d t > 0 \}, \\
t_{m}&=&\{t \mid d S(t)/d t=0 \},
\end{eqnarray}
$t_f$ denoting the firing time at which
the global variable $X(t)$ crosses the threshold
$\theta$ from below
and $t_m$ the time when $S(t)$ has the maximum value.
$S_f$ and $S_m$ 
depend on model parameters such as the noise intensities 
($\alpha$ and $\beta$), the coupling strength ($J$) 
and the size of cluster ($N$).

\section{CALCULATED RESULTS}

We have made numerical calculations, applying an external input given by
\begin{eqnarray}
I^{(e)}(t)&=& A \:\Theta(t-t_{in})\:\Theta(t_{in}+t_{w}-t),
\end{eqnarray}
where $A=0.1$, $t_{in}=40$ and $t_{w}=10$ \cite{Hasegawa03a}.
AMM equations given by Eqs. (23)-(30)
have been solved by using the fourth-order Runge-Kutta method
with a time step of 0.01.
Direct simulations (DS) for the $N$-unit FN model given by Eqs. (1)-(3)
have been performed by using the Heun method with
a time step of 0.003.
Results of DS are averaged over 100 trials.
All quantities are dimensionless.

Figures 1(a)-(d) show time courses of $\mu_1(t)$,
$\gamma_{1,1}(t)$, $\rho_{1,1}(t)$ and $S(t)$, respectively,
calculated by AMM (solid curves) and DS (dashed curves)
with $\alpha=0.01$, $\beta=0.001$, $J=1.0$ and $N=100$.
When the external input $I^{(e)}(t)$ is applied at 
$t=40$ to
the quiescent states which have been randomized by applied
small additive noises of $\beta=0.001$,  
FN neurons fire, and $\gamma_{1,1}(t)$, $\rho_{1,1}(t)$ 
and $S(t)$ develop.
Results calculated by AMM are in good agreement with those
of DS.

Figure 1(d) shows that when an input signal is applied
at $t=40$, $S(t)$ is suddenly decreased 
but has a peak at $t \sim 60$ where 
$X(t)$ is in the refractory period.
In order to see the behavior of $S(t)$ in more detail,
we show in Fig.2, its time course for the four cases:
(1) $\alpha=0.0$, (2) $\alpha=0.002$,
(3) $\alpha=0.01$ and (4) $\alpha=0.05$
with $\beta=0.001$, $J=1.0$ and $N=100$.
In the case (1), the system is subjected only to additive noise,
for which
$S(t)$ plotted by dashed curve 
is increased by an applied input for $40 \leq t < 50$.
It shows
$S_f=0.30$ at $t_f=44.5$ and $S_{m}=0.44$ at $t=60.35$, and
approaches the equilibrium value of $S=0.159$ at $t > 100$.
In the case (2) with $\alpha=0.002$,
$S(t)$ shown by dotted curve
yields $S_f=0.205$ at $t_f=44.5$ and
$S_{m}=0.526$ at $t=60.37$.
In the case (3) with $\alpha=0.01$,
$S(t)$ shown by solid curve
leads to $S_f=0.05$ at $t_f=44.5$ and
$S_{m}=0.838$ at $t=60.55$.
In the case (4) with stronger multiplicative noise of
$\alpha=0.05$, $S(t)$ plotted by chain curve
yields $S_f=0.03$ at $t_f=44.5$
and $S_m=0.910$ at $t=60.6$.
We note that with more increasing $\alpha$,
$S_f$ is much decreased while $S_{m}$ is much increased. 

This trend is more clearly seen in Fig. 3(a), where
$S_f$ and $S_{m}$ are plotted as a function of $\alpha$
for $\beta=0.001$, $0.01$ and 0.02.
In the case of $\beta=0.001$, $S_f$ is rapidly decreased and
$S_m$ is rapidly increased
with increasing $\alpha$.
For stronger additive noises of $\beta=0.01$ and $\beta=0.02$,
$S_f$ ($S_m$) is gradually decreased (increased) with increasing $\alpha$.
These results show that multiplicative noises enhance $S_m$ but
deteriorate $S_f$.

Figure 3(b) shows $S_f$ and $S_m$ 
as a function of $\beta$ for $\alpha=0.0$, $\alpha=0.01$ 
and $\alpha=0.05$.
In the case of $\alpha=0.0$, $S_f$ and $S_m$ are almost independent
of $\beta$ though they are gradually decreased for larger $\beta$.
In the cases of $\alpha=0.01$ and $\alpha=0.05$, $S_f$ and $S_m$ are 
increased and decreased, respectively,
with increasing $\beta$.
Although these results give an impression that additive noises enhance $S_f$
and deteriorate $S_m$, it is not true.
Rather additive noises work to recover $S_f$ and $S_m$
to the values of $S_f=0.30$ and $S_m=0.44$ for
the absence of multiplicative noise ($\alpha=0$).

We have studied the effects of the noises on $S_f$ and $S_m$:
the former expresses the synchronization ratio
at the firing time at $X(t)=\theta$ 
and the latter denotes its maximum value 
when $X(t)$ is in the 
refractory period at $t \sim 60$. 
We may note that if multiplicative noises exits,
the synchronization ratio $S(t)$ is once decreased when an input
applied, and it soon rebounds, showing the enhanced
value. This trend is more significant for a considerable
multiplicative noises.
In this sense, the synchronization may be enhanced by 
multiplicative noises. 

By using our AMM, it is possible to study
the dependence of the synchronization 
on the size of ensembles ($N$).
Figure 4 shows the $N$ dependences of $S_f$ and $S_m$
in the two cases for (1) $\alpha=0.01$ and $\beta=0.001$
(multiplicative noise dominant)
and (2) $\alpha=0.0$ and $\beta=0.01$
(additive noise only) with $J=1.0$.
Both $S_f$ and $S_m$ are increased with decreasing $N$,
which shows that the synchronization becomes
better for smaller systems.

\section{CONCLUSION AND DISCUSSION}


Although we have adopted the diffusive coupling 
given by Eq. (3), the sigmoid coupling given by
\begin{equation}
I_i^{(c)}(t)= \frac{K}{Z} \sum_{j\neq i} H(x_j(t)),
\end{equation}
has been widely employed for discussing networks,
where $K$ expresses a coupling strength 
and $H(x)$ is an arbitrary function of $x$ [Eq. (51)].
Diffusive and sigmoid couplings model electrical and chemical
synapses, respectively, in neuronal systems.
It is worthwhile
to study also the case of sigmoid coupling although such
FN ensembles subjected only 
to additive noises were studied
with the use of AMM \cite{Hasegawa03a}.

A straightforward calculation using the AMM 
discussed in Sec. II leads to equations of motion given by
\begin{eqnarray}
\frac{d \mu_1}{dt} &=& f_o+f_2\gamma_{1,1}-c \mu_2
+ K(h_0+h_2 \gamma_{1,1}) 
+ \frac{\alpha^2 \mu_1}{2}+I^{(e)}, \\
\frac{d \mu_2}{dt}&=& b \mu_1-d \mu_2+e, \\
\frac{d \gamma_{1,1}}{dt} &=& 2(a \gamma_{1,1}-c \gamma_{1,2})
+K h_1 (\rho_{1,1}-\frac{\gamma_{1,1}}{N}) \nonumber \\
&& +2 \alpha^2 \gamma_{1,1} + \alpha^2 \mu_1^2
+\beta^2, \\
\frac{d \gamma_{2,2}}{dt}&=& 2(b\gamma_{1,2}-d\gamma_{2,2}), \\
\frac{d \gamma_{1,2}}{dt} &=& b\gamma_{1,1}
+(a-d)\gamma_{1,2}-c\gamma_{2,2}
+Kh_1(\rho_{1,2}-\frac{\gamma_{1,2}}{N}) \nonumber  \\
&&+\frac{\alpha^2\gamma_{1,2}}{2}, \\
\frac{d \rho_{1,1}}{dt} &=& 2(a \rho_{1,1}-c \rho_{1,2})
+ 2 K h_1 \rho_{1,1}
+2 \alpha^2 \rho_{1,1} + \frac{\alpha^2 \mu_1^2}{N} 
+\frac{\beta^2}{N}, \\
\frac{d \rho_{2,2}}{dt}&=& 2(b\rho_{1,2}-d\rho_{2,2}), \\
\frac{d \rho_{1,2}}{dt} &=& b\rho_{1,1}
+(a-d)\rho_{1,2}-c\rho_{2,2}+ K h_1 \rho_{1,2}
+\frac{\alpha^2\rho_{1,2}}{2}, 
\end{eqnarray}
where $h_{\ell}=(1/\ell\: !)H^{(\ell)}(\mu_1)$.
Comparing Eqs. (43)-(50) to Eqs. (23)-(30), we note the 
following points
in coupling contributions between the two types of
couplings:
(i) the contributions in $d \gamma_{1,1}/d t$ 
and $d \gamma_{1,2}/d t$ terms
of diffusive coupling are proportional to
$(\rho_{1,1}-\gamma_{1,1})$ while those for sigmoid couplings
are proportional to $(\rho_{1,1}-\gamma_{1,1}/N)$, 
(ii) $d \mu_1/d t$, $d \rho_{1,1}/d t$ and $d \rho_{1,2}/d t$
terms in diffusive coupling
have no contributions from the couplings
in contrast to those in sigmoid coupling, and
(iii) there are no differences in $d \mu_2/d t$,
$d \gamma_{2,2}/d t$ and $\rho_{2,2}/d t$ terms.
The item (i) mainly yields the difference between the effects
of multiplicative noises on the synchronization
for the diffusive and sigmoid couplings, as will be discussed
shortly.

We have performed numerical calculations by using Eqs. (43)-(50)
with 
\begin{equation}
H(x)=\frac{1}{\{1+{\rm exp}[(x-\theta)/w] \} },
\end{equation}
$\theta$ and $w$ denoting the threshold and width, respectively. 
Time courses of $\mu_1(t)$, $\gamma_{1,1}(t)$ and $\rho_{1,1}(t)$
for $\theta=0.5$, $w=0.1$, $K=0.1$ and $N=10$ are
similar to those shown in Figs. 1 and 6 
in Ref. \cite{Hasegawa03a}.
Figure 5 shows time courses of $S(t)$ for three $\alpha$
values of $\alpha=0.0$, 0.01 and 0.05 with $\beta=0.001$.
We note that $S(t)$ has two peaks: one after an input is
applied and the other when $X(t)$ is in the refractory period.
In the case of $\alpha=0.0$, we get $S_f=0.108$ at $t=44.16$
and $S_m=0.342$ at $t=62.92$.
In the case of $\alpha=0.01$, we get $S_f=0.073$ at $t=44.16$
and $S_m=0.287$ at $t=64.35$.
In the case of $\alpha=0.05$, we get $S_f=0.053$ at $t=44.15$
and $S_m=0.284$ at $t=64.32$.
Both $S_f$ and $S_m$ are decreased with increasing $\alpha$.
This behavior is more clearly shown in Fig. 6
where $S_f$ and $S_m$ are plotted as a function of $\alpha$.

A comparison between Figs. 3(a) and 6 shows that
with increasing $\alpha$,
$S_m$ for diffusive couplings is increased
while that for sigmoid couplings is decreased.
This difference mainly arises
from the item (i) for the $d \gamma_{1,1}/d t$ term
discussed above. Namely,
$\gamma_{1,1}$ for diffusive coupling
is reduced by a {\it negative} contribution
proportional to $(\rho_{1,1}-\gamma_{1,1})$ 
({\it i.e.} $\rho_{1,1} < \gamma_{1,1}$)
which yields an enhancement in $S(t)$ of Eq. (36).
On the contrary, 
$\gamma_{1,1}$ for sigmoid coupling
is slightly increased by a {\it positive} contribution
proportional to $(\rho_{1,1}-\gamma_{1,1}/N)$ 
({\it i.e.} $\rho_{1,1} > \gamma_{1,1}/N$)
which reduces $S(t)$ of Eq. (36).
In contrast,
effects of multiplicative noise 
are not effective for $S_f$ because $x_i$ 
is not large at $t_{in} < t < t_f$.
Then with increasing $\alpha$,
$S_f$ is decreased for both the couplings. 
The main difference between the two couplings is
the presence of the feedback (second) term of Eq. (3).
Indeed, if we adopt the diffusive coupling
without this term, which is equivalent to 
the sigmoid coupling with $H(x)=x$ in Eq. (42),
the synchronization is decreased with increasing $\alpha$
(result not shown).
This situation is similar to that
of the synchronization in small-world networks \cite{Wat98}.
It was shown in Ref. [15e] that when
a small-world network is made by introducing randomness
to a regular network, the synchronization in the small-world
network with diffusive couplings 
may be increased while that with sigmoid couplings 
is decreased.
Our calculation implies that
synchronization depends not only on the type of noises
but also on the kind of couplings.
This may also suggest that an ordered state
in the multiplicative noise-induced phase transition  
reported in Refs. \cite{Broeck94}-\cite{Munoz05},
might partly owe the diffusive couplings
employed in these studies: multiplicative noises
could not yield an ordered state with sigmoid couplings .

In summary, we have studied the synchronization
in FitzHugh-Nagumo neuronal ensembles subjected to
additive and multiplicative noises, by reformulating
AMM with the use of FPE \cite{Hasegawa06,Hasegawa03a}.
The property of the two noises in FN neuron ensembles
is summarized as follows. 

\noindent
(a) the scaling relation: $\alpha(N)=\alpha(1)/\sqrt{N}$
is not hold for multiplicative noises although the relation:
$\beta(N)=\beta(1)/\sqrt{N}$ is valid for additive noises,

\noindent
(b) multiplicative noises may enhance the synchronization ($S_m$)
for diffusive couplings
though both the two noises are generally detrimental to it, and

\noindent
(c) for both the additive and multiplicative noises, 
the synchronization is more increased
in smaller $N$ systems.

\noindent
The item (a) supplements the result for Langevin model
\cite{Hasegawa06}.
The item (b) is similar to the item (1) of
an ordered state created by multiplicative noises 
\cite{Broeck94}-\cite{Munoz05} mentioned in the introduction.

A disadvantage of our AMM is that its applicability 
is limited to weak-noise cases. 
For multiplicative Gaussian noises, the probability distribution 
become non-Gaussian yielding divergent
second and higher moments for a large $\alpha$, to which
the AMM cannon be applied.
On the contrary, an advantage of the AMM is that we can easily discuss
dynamical property of the finite $N$-unit stochastic systems.
We have solved the eight-dimensional ordinary differential equations
for FitzHugh-Nagumo neuronal ensembles.
In contrast, within direct simulation and the FPE approach,
we have to solve
the $2 N$-dimensional stochastic equations and
the $(2 N+1)$-dimensional partial differential 
equations, respectively,
which are much laborious than AMM.
Our AMM may be applied to a wide class of coupled stochastic models
subjected to additive and/or multiplicative noises.

\section*{Acknowledgements}
This work is partly supported by
a Grant-in-Aid for Scientific Research from the Japanese 
Ministry of Education, Culture, Sports, Science and Technology.  


\vspace{1cm}
\noindent
{\Large\bf Appendix: Equations of motions for a general $G(x)$}

Although Eqs. (23)-(30) express equations 
of motion for $G(x)=x$, we present the result for a
general form of $G(x)$:
\begin{eqnarray}
\frac{d \mu_1}{dt} &=& f_o+f_2\gamma_{1,1}-c \mu_2
+\frac{\alpha^2}{2}[g_0g_1+3(g_1g_2+g_0g_2)\gamma_{1,1}]
+I^{(e)}, \\
\frac{d \mu_2}{dt}&=& b \mu_1-d \mu_2+e, \\
\frac{d \gamma_{1,1}}{dt} &=& 2(a \gamma_{1,1}-c \gamma_{1,2})
+\frac{2w N}{Z} (\rho_{1,1}-\gamma_{1,1})
+2\alpha^2 (g_1^2+2g_0g_2) \gamma_{1,1} \nonumber \\
&+& \alpha^2 g_0^2
+\beta^2, \\
\frac{d \gamma_{2,2}}{dt}&=& 2(b\gamma_{1,2}-d\gamma_{2,2}), \\
\frac{d \gamma_{1,2}}{dt} &=& b\gamma_{1,1}
+(a-d)\gamma_{1,2}-c\gamma_{2,2}
+\frac{w N}{Z}(\rho_{1,2}-\gamma_{1,2})
+\frac{\alpha^2}{2}(g_1^2+2g_0g_2)\gamma_{1,1}, \\
\frac{d \rho_{1,1}}{dt} &=& 2(a \rho_{1,1}-c \rho_{1,2})
+2\alpha^2 (g_1^2+2g_0g_2)  \rho_{1,1} 
+ \frac{\alpha^2 g_0^2}{N} 
+\frac{\beta^2}{N}, \\
\frac{d \rho_{2,2}}{dt}&=& 2(b\rho_{1,2}-d\rho_{2,2}), \\
\frac{d \rho_{1,2}}{dt} &=& b\rho_{1,1}
+(a-d)\rho_{1,2}-c\rho_{2,2}
+\frac{\alpha^2}{2}(g_1^2+2g_0g_2)\rho_{1,2}, 
\end{eqnarray}
where 
$g_{\ell}=(1/\ell \:!)G^{(\ell)}(\mu_1)$.
For $G(x)=x$, we get $g_0=\mu_1$, $g_1=1$, and $g_2=g_3=0$,
with which Eqs. (52)-(59) reduce to Eqs. (23)-(30).

\newpage


\newpage

\begin{figure}
\caption{
(color online).
Time courses of (a) $\mu_1(t)$, (b) $\gamma_{1,1}(t)$, 
(c) $\rho_{1,1}(t)$ and (d) $S(t)$
for $\alpha=0.01$, $\beta=0.001$ with
diffusive coupling (DC) of $J=1.0$ and $N=100$,
solid and dashed curves denoting results of AMM
and direct simulations, respectively.
At the bottom of (a), an input signal is plotted. 
Vertical scales of (b) and (c) are multiplied
by factors of $10^{-4}$ and $10^{-5}$,
respectively.
}
\label{fig1}
\end{figure}

\begin{figure}
\caption{
(color online).
Time courses of the synchronization ratio $S(t)$
for $\alpha=0.0$ (dashed curve), $\alpha=0.002$ (dotted curve),
$\alpha=0.01$ (solid curve)
and $\alpha=0.05$ (chain curve) with
diffusive couplings of $J=1.0$, $\beta=0.001$ and $N=100$.
}
\label{fig2}
\end{figure}

\begin{figure}
\caption{
(color online).
(a) $\alpha$ dependences of $S_f$ and $S_m$
for $\beta=0.001$ (circles), $\beta=0.01$ (squares)
and $\beta=0.02$ (triangles), and
(b) $\beta$ dependences of $S_f$ and $S_m$
for $\alpha=0.0$ (circles), $\alpha=0.01$ (squares)
and $\alpha=0.05$ (triangles)
with diffusive coupling of $J=1.0$ and $N=100$:
results of $S_m$ and $S_a$ of AMM are expressed by
solid and chain curves, respectively, and those of DS by
filled and open marks, respectively.
}
\label{fig3}
\end{figure}

\begin{figure}
\caption{
(color online).
$N$ dependences of $S_f$ (solid curve) 
and $S_{m}$ (chain curve)
for two sets of parameters:
(1) $\alpha=0.01$ and $\beta=0.001$ and 
(2) $\alpha=0.0$ and $\beta=0.01$ 
with diffusive coupling of $J=1.0$
and $N=100$, calculated by AMM
and DS (circles and squares).
}
\label{fig4}
\end{figure}

\begin{figure}
\caption{
(color online).
Time courses of the synchronization ratio $S(t)$
for $\alpha=0.0$ (dashed curve), 
$\alpha=0.01$ (solid curve)
and $\alpha=0.05$ (chain curve)
with sigmoid couplings (SC) of $K=0.1$, $\beta=0.001$ and $N=10$.
}
\label{fig5}
\end{figure}

\begin{figure}
\caption{
(color online).
$\alpha$ dependences of $S_f$ and $S_m$
for $\beta=0.001$ (circles) and $\beta=0.01$ (squares)
with sigmoid coupling of $K=0.1$ and $N=10$:
results of $S_m$ and $S_a$ of AMM are expressed by
solid and chain curves, respectively, and those of DS by
filled and open marks, respectively.
}
\label{fig6}
\end{figure}

\end{document}